\documentclass[aps,pra,superscriptaddress,twocolumn,nofootonbib,floatfix,longbibliography]{revtex4-2} 

\usepackage[final]{graphicx}    
\graphicspath{{figures/}}       

\usepackage{dcolumn} 
\usepackage{bm} 
\usepackage{amsmath}
\usepackage{amsfonts} 
\usepackage{amssymb}
\usepackage{bbold}
\usepackage[utf8]{inputenc}
\usepackage{textcomp}
\usepackage[english]{babel}
\usepackage[usenames,dvipsnames]{xcolor}
\usepackage{courier}
\usepackage{hyperref}
\hypersetup{colorlinks=true}
\usepackage{listings}
\usepackage{color}

\definecolor{dkgreen}{rgb}{0,0.6,0}
\definecolor{gray}{rgb}{0.5,0.5,0.5}
\definecolor{mauve}{rgb}{0.58,0,0.82}

\begin{document}
\title{Statistical complexity and the road to equilibrium in many-body chaotic quantum systems}

\author{Manuel H. Muñoz-Arias}
\email{mhmunoz@unm.edu}
\affiliation{Center for Quantum Information and Control, CQuIC, Department of 
Physics and Astronomy, University of New Mexico, Albuquerque, New Mexico 
87131, USA}

\begin{abstract}
In this work we revisit the problem of equilibration in isolated many-body interacting quantum systems. We pay particular attention to quantum chaotic Hamiltonians, and rather than focusing on the properties of the asymptotic states and how they adhere to the predictions of the Eigenstate Thermalization Hypothesis, we focus on the equilibration process itself, \textit{i.e.}, \emph{the road to equilibrium}. Along the road to equilibrium the diagonal ensembles obey an emergent form of the second law of thermodynamics and we provide an information theoretic proof of this fact. With this proof at hand we show that the road to equilibrium is nothing but a hierarchy in time of diagonal ensembles. Furthermore, introducing the notions of statistical complexity and the entropy-complexity plane, we investigate the uniqueness of the road to equilibrium in a generic many-body system by comparing its trajectories in the entropy-complexity plane to those generated by a random Hamiltonian. Finally by treating the random Hamiltonian as a perturbation we analyzed the stability of entropy-complexity trajectories associated with the road to equilibrium for a chaotic Hamiltonian and different types of initial states.  
\end{abstract}

\date{\today}

\maketitle

\section{Introduction}
\label{sec:intro}
Strongly interacting quantum many-body systems far from equilibrium~\cite{Polkovnikov2011} have helped us to establish a modern interpretation and understanding of some traditional topics in quantum theory, as for instance quantum chaos~\cite{Roberts2017,Wimberger2020}. At the same time, their study has implications in a variety of other fields. For example, quantum simulation and computation~\cite{Heyl2019err,Sieberer2019}, and more recently, quantum error correction with the advent of the so called Floquet codes~\cite{Hastings2021dynamically}. Their study has also helped in the establishment of new research topics with a wide range of application. Relevant examples are, the notion of quantum information scrambling~\cite{Sachdev1993,Maldacena2016,Swingle2016,Swingle2018}, dynamical quantum phase transitions~\cite{heyl2018} and dynamical criticality~\cite{Heyl2013} and more recently, out-of-equilibrium phases of matter~\cite{Else2020,Haldar2021}.

Importantly, a foundation for the emergence of statistical mechanics in closed quantum systems has resulted from investigations of the nature and structure of Hamiltonians describing strongly interacting many-body quantum systems. This result has the form of the celebrated, and now widely accepted, eigenstate thermalization hypotesis (ETH)~\cite{Srednicki1994,Deutsch1991,Rigol2008,Deutsch2018}. In a simplified form it states that eigenstates in the bulk of the many-body spectrum are their own microcanonical ensembles, as such, expectation values of observables can be readily computed by the microcanonical average over a small energy window centered at the mean energy of the initial state.

Despite all these different advances, the physical nature of the mechanism responsible for the equilibration of many-body closed quantum systems under their own dynamics~\cite{Polkovnikov2011,Rigol2016}, in a generic many-body quantum system, is still object of ongoing work~\cite{Eisert2015,Gogolin2016,Langen2015,Rigol2008,Wilming2019, Balz2016,Oliveira2018,Torres2014,Rigol2016,Brenes2020}. A rather widely accepted point of view is that interaction terms, which yield the Hamiltonian nonintegrable, or even more, quantum chaotic, will generate time evolution leading to equilibration for almost all states beginning out of equilibrium.

One method to bring interacting Hamiltonian terms into the picture is a Hamiltonian quench~\cite{Polkovnikov2011}. One considers a time-independent Hamiltonian with a control parameter, and picks the initial state to be a stationary state of the Hamiltonian for a given value of the control parameter. Then suddenly changing the value of the control parameter such that the initial state will begin out-of-equilibrium. In this work we study the nonequilibrium dynamics of a chaotic many-body quantum system after a Hamiltonian quench. We focus on the time scale between $t=0$ and $t = t_{\rm equiv}$ the equilibration time, \textit{i.e.}, the transient dynamics or local equilibration, which we refer to as the \emph{road to equilibrium}. In contrast with the more estandarized studies of equilibrium properties, where observable quantities can be computed using the appropriate random matrix ensemble depending on the system symmetries. Usually the Gaussian Orthogonal ensemble (GOE) or Circular Orthogonal Ensemble (COE), for Hamiltonians or unitary operators which have time reversal symmetry. By making use of the notion of statistical complexity~\cite{Crutchfield1989,Crutchfield1989,Feldman1997,Crutchfield2012}, a common tool in time series analysis, we will show that the road to equilibrium is highly nontrivial and markedly different to that obtained from evolution under a random Hamiltonian, which nevertheless, captures almost perfectly the system properties once equilibrium has been reached. 

As an example of a closed many-body quantum system, which is known to be 
quantum chaotic~\cite{Kolovsky2004a,Cruz2020}, we consider the single-band 
Bose-Hubbard (BH) model and will pay special attention to the time evolution 
of states in the Fock (site occupation number) basis, and study the trajectories of the transient dynamics in the entropy-complexity plane~\cite{Martin2006,Rosso2007}. Nonequilibrium dynamics of this model has been studied extensively, both  theoretically~\cite{Horiguchi2009,Bloch2008,Kollath2007,Trefzger2011, Shimizu2018,Sorg2014} and experimentally~\cite{Ronzheimer2014}, using ultracold bosonic atoms trapped in optical lattices.

The rest of this manuscript is organized as follows. In Sec.~\ref{sec:equilibration_in_many_body} we discuss the idea of the road to equilibrium in many-body chaotic quantum systems and its connection to an emergent second law of thermodynamics, and as a complement we present an information theoretic proof of the emergence of a notion of second law. In  Sec.~\ref{sec:st_and_sc_plane} we introduce the concept of statistical complexity and the way to quantify it. At the same time we introduce a way of 
visualizing the dynamics of the statistical complexity known as the 
entropy-complexity (E-C) plane. In Sec.~\ref{sec:BH_model} we briefly comment 
on the BH model, which will be our working example of quantum chaotic 
many-body system. Results on the dynamics of  statistical complexity during equilibration and its implications for the  physical characterization of the road to equilibrium are presented in Sec.~\ref{sec:preliminary_results}. To complement these results we include, in Sec.~\ref{sec:preliminary_results}, a study of the effects on the road to equilibrium of a random perturbation, in particular a clear separation between coherent quantum dynamics and random evolution is seen in the E-C plane. A summary and outlook is presented in Sec.~\ref{sec:final_remarks}.

\section{The road to equilibrium in closed many-body chaotic quantum systems and the second law of thermodynamics}
\label{sec:equilibration_in_many_body}
We consider equilibration in a closed many-body quantum system as consequence of a quantum quench. This scheme begins with an initial state which is ``easy'' to prepare, that we assume to be an eigenstate of the Hamiltonian prior to the quench. In the case of a spin system one might prepare a spin coherent state, for instance the stretched state $|\uparrow\rangle^{\otimes N}$, where $N$ is the number of spins in the chain, and $\hat{\sigma}_z|\uparrow\rangle = |\uparrow\rangle$ with $\hat{\sigma}_z$ the $z$ Pauli matrix. In the case of a bosonic lattice one might prepare a state with well define site occupations $|n_1\rangle\otimes...\otimes|n_L\rangle$, where $L$ is the number of sites on the lattice, and $\hat{n}_j|n_j\rangle = n_j |n_j\rangle$ with $\hat{n}_j$ the number operator for site $j$. Assuming the the initial state is an eigenstate of the Hamiltonian prior to the quench already fixes a preferred basis. In the above examples we recognize the computational basis in the case of the spin chain, and the Fock (site occupation number) basis in the case of the bosonic lattice. Once the initial state has been prepared, the Hamiltonian is quenched, this action drives the initial state out-of-equilibrium, guaranteeing that a nontrivial time evolution will follow and eventually, after a time $t_{\rm equiv}$, equilibrium will be reached. 

Given our choice of reference basis, we can study the road to equilibrium by investigating the properties of the states from $t=0$ up to $t_{\rm equiv}$. Described in this basis, the road to equilibrium is an nonequilibrium process with fixed points on both ends. On the one hand, the initial state gives rise to a probability distribution assigning full certainty to a single event, \textit{i.e.}, observation of the initial state. On the other hand, after equilibrium has been reached, expectation values of observables agree with the values computed from one of the ensembles of equilibrium statistical mechanics~\footnote{With the exception of integrable systems where generalized Gibbs ensembles are the ones capturing the system properties at asymptotic times~\cite{Rigol2007,caux2011}}. As such, the statistical ensemble provides an acceptable representation of the state of the system.

Notice that the road to equilibrium takes a simple initial state, assigning complete certainty to a single event, into a ``delocalized'' state, making an assignment of probability weights to a large set of events. This implies that as the system evolves towards equilibrium, knowledge of the initial state is lost, and thus entropy increases. In a classical thermodynamic setting, where system is coupled to a large thermal bath, the above statement is summarized in the form of the second law of thermodynamics, guaranteeing the monotony of the thermodynamic entropy. However, as we are working with a closed quantum system, there is no notion of temperature in the usual sense, and thus no clear notion of thermodynamic entropy. An alternative is to use the diagonal entropy~\cite{Rigol2016}, which is constructed as the Shannon information entropy of the diagonal ensemble, the latter is given by the diagonal entries of the density matrix $\rho$, that is
\begin{equation}
 \label{eqn:diag_ensemble}
 \rho|_{\rm diag} = \sum_{m=1}^{\mathcal{D}_\mathcal{H}} \rho_{mm}|m\rangle\langle m|,
\end{equation}
where $\{|m\rangle\}$ is a reference basis~\footnote{Which we will fix to be the eigenbasis of the Hamiltonian prior to the quench.}, $\rho_{mm}$ are the diagonal entries of the density matrix in the reference basis, and $\mathcal{D}_\mathcal{H}$ is the Hilbert space dimension. With this definition of the diagonal ensemble we write the diagonal entropy as
\begin{equation}
\label{eqn:diag_entropy}
 S_{\rm diag} = -\sum_{n=1}^{\mathcal{D}_\mathcal{H}} \rho_{nn}\ln(\rho_{nn}),
\end{equation}
that is, the Shannon entropy of the diagonal entries of $\rho$ in the reference basis. It can be shown~\cite{Rigol2016} that for quantum chaotic systems the diagonal entropy in Eq.~(\ref{eqn:diag_entropy}) is a good thermodynamic entropy, and thus the second law of thermodynamics determines its nonequilibrium evolution.

We then reach the following statement: the road to equilibrium in a closed many-body interacting chaotic quantum system can be understood as thermal, via the diagonal entropy, and as such, it is govern by an emergent form of a second law of thermodynamics. This has been put into formal footing in Refs.~\cite{Sorg2014,Ikeda2015,Rigol2016}. In other words, if $t<t'$ then it is true that
\begin{equation}
S_{\rm diag}(t) < S_{\rm diag}(t'),
\end{equation}
under unitary evolution for a closed interacting many-body chaotic Hamiltonian. The first contribution of this work is to present an alternative proof of this fact based on tools from the theory of majorization, given in the next subsection.

\subsection{An information theoretic view on the second law of thermodynamics for closed quantum systems}
\label{sec:second_law}
In order to show that $S_{\rm diag}$ obeys a second law of thermodynamics, we need to show two things. First, that $S_{\rm diag}$ grows monotonically with time as a consequence of the nonequilibrium dynamics, and second that it is a good thermodynamic entropy.

Monotonicity can be shown using tools from the theory of \textit{majorization}~\cite{Ando1989,Alberti1982,Bhatia1997,Marshall2011}. We introduce the necessary ones here. Consider two positive real $\mathcal{D}$-dimensional vectors, $\vec{p}$ and $\vec{q}$. We say that $\vec{p}$ majorizes $\vec{q}$ if
\begin{equation}
 \label{eqn:majorization_relation}
  \sum_{l=1}^{k}\vec{p}_l^\downarrow \ge \sum_{l=1}^k\vec{q}_l^\downarrow
\end{equation}
for all $k = 1,...,\mathcal{D}-1$, with strict equality when $k=\mathcal{D}$, and write $\vec{p} \succ \vec{q}$. The notation $^\downarrow$ indicates that the vector is sorted in decreasing order. The definition in Eq.~(\ref{eqn:majorization_relation}) does not talk about the normalization of the vectors, and thus it applies to probability vectors, where $\sum_{n=1}^\mathcal{D}\vec{p}_n = 1$, as the ones we are concerned about for this work. The relation $\succ$ induces a partial order in the space of probability vectors, since there exist pairs of vectors where neither one majorizes the other. Intuitively we can understand the majorization relation 
$\vec{p} \succ \vec{q}$ as telling us that $\vec{p}$ is a narrower distribution, hence less uncertain than $\vec{q}$.

The key result from the theory of majorization which allows us to show the monotonicity of $S_{\rm diag}$, is the connection between majorization and Schur-concave functions~\cite{Marshall2011}. It is known that if
\begin{equation}
 \label{eqn:majorization_concave}
 \vec{p} \succ \vec{q} \Rightarrow \mathcal{F}[\vec{p}]\ge \mathcal{F}[\vec{q}],
\end{equation}
where $\mathcal{F}$ is any Schur-concave function of the probability vector.

In order to use this statement we need to show that the time evolution of our closed quantum system generates a hierarchical structure, based on the relation $\succ$, of the diagonal ensembles which is strictly descending. This implies the diagonal ensemble at any new time instant must always be majorized by those diagonal ensembles at previous time instants. This fact follows from Horn's Lemma~\cite{Horn1954,Ando1989}. Given $\vec{p}$ and 
$\vec{q}$ two $\mathcal{D}$-dimensional vectors, then 
\begin{equation}
\label{eqn:Horn_lemma}
 \vec{p}\succ\vec{q} \Longleftrightarrow \vec{q} = D\vec{p},
\end{equation}
where $D$ is a unitary-stochastic matrix. An elegant proof of this statement, other than the original, was given by Nielsen in~\cite{Nielsen2000}.

In order to see how time evolution in our isolated system can be related to Eq.~(\ref{eqn:majorization_concave}) and Eq.~(\ref{eqn:Horn_lemma}), we follow~\cite{Rigol2016}. Let us now go back to the physical situation under 
investigation. Initially, we prepare our system in a stationary state, which then undergoes unitary evolution in response to an external change. The external change induces a change in the Hamiltonian parameters, taking place during a finite time interval (instantaneous in the case of quench dynamics), after which the Hamiltonian is again time independent.

From the point of view of the state, even though we began with a stationary state (diagonal $\rho$ in the basis of the initial Hamiltonian), after the external process, the state is not stationary anymore. Yet, under the lens of ETH, expectation values of observable quantities are completely determined by the diagonal entries of $\rho$, thus we can write the unitary evolution as 
\begin{equation}
\label{eqn:diago_evol}
 \rho_{m'm'} = \sum_{m=1}^\mathcal{D}U_{m'm}\rho_{mm}^{(0)}U_{m'm}^\dagger,
\end{equation}
where $\{|m'\rangle\}$ is the eigenbasis of the initial Hamiltonian, $\{|m\rangle\}$ is the eigenbasis of the Hamiltonian after the external process, and $U_{m'm} = \langle m' |\hat{U}|m \rangle$, with $\hat{U}$ the time 
evolution operator. Eq. (\ref{eqn:diago_evol}) can be written as
\begin{equation}
\label{eqn:diago_evol_probs}
 \rho_{m'm'} = \sum_{m=1}^\mathcal{D}\rho_{mm}^{(0)}P_{m\to m'},
\end{equation}
where $P_{m\to m'} = U_{m'm}U_{mm'}^\dagger = |U_{m'm}|^2$. Let us explore further the nature of these transition probabilities. From the unitarity of the time evolution we know 
\begin{equation}
 \sum_{m'}U_{m'n}U_{sm'}^\dagger = \delta_{ns}, \quad 
\sum_{n}U_{nk'}^\dagger U_{m'n} = \delta_{k'm'},
\end{equation}
thus 
\begin{eqnarray}
 \label{eqn:stochastic_condition_1}
 \sum_{m'}U_{m'n}U_{nm'}^\dagger = \sum_{m'}P_{n\to m'} = 1, \\
 \label{eqn:stochastic_condition_2}
 \sum_{n}U_{nm'}^\dagger U_{m'n} = \sum_n P_{n\to m'} = 1.
\end{eqnarray}
Noticing that all the $P_{n\to m'}$ are always positive, we have 
\begin{equation}
 \label{eqn:stochastic_condition_3}
 0 \le P_{n\to m'} \le 1.
\end{equation}
Any semi-positive matrix $\mathbb{P}$ satisfying Eq.~(\ref{eqn:stochastic_condition_1}), Eq.~(\ref{eqn:stochastic_condition_2}), and Eq.~(\ref{eqn:stochastic_condition_3}), is called doubly stochastic. Furthermore, the matrix with entries $\mathbb{P}_{nm'} = P_{n\to m'} = |U_{nm'}|^2$ is called unitary-stochastic.

We see then that, assuming the stationarity of the initial density matrix and invoking ETH, $\rho|_{\rm diag}$ evolves following Eq.~(\ref{eqn:diago_evol_probs}), which is an evolution driven by a unitary stochastic matrix. Therefore, Horn's lemma, Eq.~(\ref{eqn:Horn_lemma}), implies that $\rho|_{\rm diag}(t) \succ \rho|_{\rm diag}(t')$ for any $t'>t$. Then, Eq.~(\ref{eqn:majorization_concave}) applied to $S_{\rm diag}$ implies 
\begin{equation}
 S_{\rm diag}(t') \ge S_{\rm diag}(t),
\end{equation}
for all $t' \ge t$. Which completes the proof of the monotonicity of the diagonal entropy.

Other consequence of the unitary-stochastic character of time evolution is the fact that, in the long time limit, any initial distribution tends towards the eigenvector of eigenvalue $1$. This eigenvector acts as an attractor, and is of the form $\rho_{\infty} = \frac{1}{\mathcal{D}}\mathbb{1}$, \textit{i.e.}, the ``infinite temperature''. As such, $S_{\rm diag}$ is bounded from above by $S_{\infty} = \ln(\mathcal{D})$, and for ant $t \le t'$ one has $S_{\rm diag}(t) \le S_{\rm diag}(t') < \ln(\mathcal{D})$.

After showing the monotonous character of the diagonal entropy, we need to show it is a good thermodynamic entropy, possibly up to subextensive fluctuations. A detailed proof of this fact can be found in chapter $5$ of~\cite{Rigol2016}. Here we ill mention that, if one specializes to quantum chaotic Hamiltonians, one can guarantee the thermodynamic character of the diagonal entropy up to subextensive fluctuations~\cite{Rigol2016}. This is why from now on we will only be concern with many-body quantum systems which are quantum chaotic.

Motivated by this complementary interpretation of the road to equilibrium in many-body interacting chaotic quantum systems after a quantum quench, in the rest of this work we will employ the tools of statistical complexity and the entropy-complexity plane to present a more in-depth characterization of the hierarchy of diagonal ensembles composing the road to equilibrium. We initiate the presentation of these results with a brief introduction to the notions of statistical complexity and the entropy-complexity plane.

\section{Statistical complexity and the entropy-complexity plane}
\label{sec:st_and_sc_plane}
The notion of complexity is perhaps one of the most attractive ideas in the natural sciences. On the one hand, it fundamentally appeals to physical intuition, as based on common sense it is really easy to say whether a physical process is complex or not. On the other hand, its proper and precise quantification is extremely challenging, and has been the topic of extensive research work in the past couple of decades.  

Due to this innate difficulty in many cases, rather than presenting a universally valid notion of complexity, narrowing down the set of physical processes whose complexity needs to be quantified, helps in simplifying this challenging tasks. In this spirit, different notions of complexity have emerged in the past, coming from a variety of subdisciplines. For instance, one can quantify complexity via the length of the minimum program needed to generate certain output bit string, which is known as the Kolmogorov complexity~\cite{Kolmogorov1998}. In a system composed of several different individual constituents one can look for the degree of cooperativity or collective phenomena~\cite{Crutchfield1994,Koppel1987}, or use tools from the domain of differential geometry to construct complexity quantifiers~\cite{Felice2018}, computational complexity and the complexity classes~\cite{Arora2009computational}, among several others~\cite{Gell1996,Badii1997}.

As discussed in Sec.~\ref{sec:equilibration_in_many_body}, the road to equilibrium is governed by an emergent second law of thermodynamics. In terms of the diagonal ensembles, it is manifested as a hierarchy of them in time. In other words, we have simple distributions describing the states on both ends of the equilibration dynamics, as such, if the system access any complex states during the road to equilibrium, this must take place at intermediate times, and it can be studied via their statistical complexity~\cite{Feldman1997,Crutchfield2012}. 

In a general context, the evolution of the diagonal ensemble is doubly stochastic (see Sec.~\ref{sec:second_law}), implying that equilibrium should be characterized by $\rho_\infty = \frac{1}{\mathcal{D}}\mathbb{1}$ or a nearby distribution $\rho_\infty + \delta\rho$ with $\delta\rho$ a small increment. This limit distribution is featureless and it tends to a uniform distribution when $\delta\rho\to0$. Thus a natural way of quantifying the complexity of a given diagonal ensemble is to take $\rho_{\infty}$ as a reference, and compute how far a given diagonal ensemble is from the equilibrium distribution, \textit{i.e.}, by defining a disequilibrium quantifier. The latter is a measure constructed via a divergence (or relative entropy) between the distribution under study and $\rho_{\infty}$. This way of quantifying complexity is referred to as statistical complexity, and it was first introduced and characterized in~\cite{Lopez1995,Feldman1997} and later extended in~\cite{Martin2006,Rosso2007}~\footnote{Other ways of quantifying statistical complexity have been defined and used in different contexts, see for instance~\cite{Feldman2008}. A popular one which was recently adapted to fit with the quantum formalism uses $\epsilon$-machines~\cite{Tan2014}.}, and it is the notion of statistical complexity we will use in the setup under investigation in this manuscript.

We define the disequilibrium quantifier as 
\begin{equation}
 \label{eqn:disequilibrium}
 \mathcal{Q}[P] = \mathcal{Q}_0 D[P, P_e],	 
\end{equation}
where $\mathcal{Q}_0$ is a normalization factor chosen such that $0 \le \mathcal{Q} \le 1$, $P_e$ is the uniform distribution or $\rho_{\infty}$, and $D$ is a measure of distance in probability space or divergence. Based on the choice of $D[.,.]$ one can define a variety of disequilibrium quantifiers, a thorough list of possibilities with their respective normalization factors is given in~\cite{Martin2006}. For the results presented in this work we will consider the Jensen-Shannon divergence~\cite{Nielsen2010,Nielsen2019}, a symmetrized version of the well known Kullback-Lieber divergence~\cite{Kullback1951}, which satisfies the triangle inequality and hence constitutes a true metric~\footnote{Recently the metric property of the quantum Jensen-Shannon divergence was analyzed~\cite{Virosztek2019}, which should, in principle, allow us to construct a full quantum statistical complexity quantifier.}. The Jensen-Shannon divergence is give by
\begin{equation}
\label{eqn:jensen_shannon}
D[P, P_e] = S\left[\frac{P+P_e}{2} \right] - \frac{1}{2}S[P] - 
\frac{1}{2}\ln(\mathcal{D}),
\end{equation}
where $S[P]$ is the Shannon information entropy of the distribution $P$. 

Using the Eq.~(\ref{eqn:disequilibrium}) we can quantify how far a distribution is from the featureless distribution, however a distribution assigning complete certainty to a single event, say an eigenstate of 
the initial Hamiltonian written in its eigenbasis, will have a large value of $D$ in Eq.~(\ref{eqn:disequilibrium}). This type of distributions can be seen as ``featureless'' and should be characterized by a zero statistical complexity. To achieve this we define the statistical complexity quantifier $C[.]$ as the product
\begin{equation}
\label{eqn:statistical_comple}
 C[P] = \mathcal{Q}[P]H[P],
\end{equation}
where $H[P]$ is a normalized measure of information or uncertainty (entropy). As with $\mathcal{Q}$, one has different choices for $H$. For instance Renyi, Tsallis, or Shannon entropies are good candidates~\cite{Martin2006}. For the present study we will consider the normalized Shannon entropy $H[P] = S[P]/\ln(\mathcal{D})$. However, the results on the dynamics of $C$ as presented in Sec.~\ref{sec:preliminary_results} do not depend on the choice of 
$\mathcal{Q}$ or $H$. The definition of statistical complexity in 
Eq.~(\ref{eqn:statistical_comple}) satisfies two constraints, it vanishes
at complete certainty (eigenstate of the initial Hamiltonian) and it 
vanishes at complete uncertainty $H[P] = 1$ (equilibrium state $\rho_{\infty}$). Thus it provides a way of quantifying the structures arising in all the diagonal states the system occupied along the road to equilibrium having these two states as boundary points.

The monotonic character of $S_{\rm diag}$ allow us to use $H[P]\in[0,1]$ as a time dimension, and thus provides a way of studying the dynamics of statistical complexity which involves only the quantities in Eq.~(\ref{eqn:statistical_comple}). We notice that $C[.]$ is not a trivial function of $H[.]$, and thus for a given value of $H$, the possible values of $C$ are bounded from above and below. That is, $C[P]$ yields a value within a range $\left[ C[P_{\rm lower}], C[P_{\rm upper}]\right]$. The region of the $(H,C)$ space in between these two boundary curves is known as the entropy-complexity (E-C) plane~\cite{Rosso2007}. By looking at the trajectories in this plane we can build an intuition of the behavior of the statistical complexity during 
equilibration dynamics of a many-body quantum system. The explicit form of the distributions saturating the lower and upper bound curves of the E-C plane are of great interest to us. Those saturating the lower bound tell us what type of distributions are seen as having the less complex structures by $C$, and those saturating the upper bound curve will give the ones of maximum $C$, for all values of $H\in[0,1]$. We present the details of the construction of these ``boundary'' distributions in App.~\ref{sec:app_boundaries}.

\section{The single-band Bose-Hubbard model}
\label{sec:BH_model}
As discussed in Sec.~\ref{sec:intro}, the quantum chaotic many-body model we consider to investigate the behavior of the statistical complexity during equilibration dynamics is the single-band Bose-Hubbard model. Originally proposed in~\cite{Fisher1989}, and later on studied in~\cite{Jaksch1998} as a model for quantum computation, the BH Hamiltonian describes identical bosonic particles trapped in a periodic potential. Individual particles are allowed to hop between lattice sites with a hopping amplitude $J$, and we will restrict ourselves to the case of nearest neighbor hopping. Additionally, doubly or higher occupations of a single lattice site are penalized with an energy cost proportional to $U$ (the particle-particle interaction strength). Several out-of-equilibrium aspects of this model have been extensively studied in the past couple of decades, with particular emphasis on their connections to experiments with ultracold atoms in optical lattices~\cite{Bloch2008}.

In the present work we will restrict ourselves to the single energy band limit in one spatial dimension. Under these constraints the BH Hamiltonian reads
\begin{equation}
 \label{eqn:BH_hamiltonian}
 \hat{H}_{\rm BH} = \hat{H}_{\rm hop} + \hat{H}_{\rm int},
\end{equation}
where $\hat{H}_{\rm hop}$ represents the kinetic energy part of the Hamiltonian, accounting for hopping events, and $\hat{H}_{\rm int}$ represents the interaction energy, due to particle-particle interactions via collisions. These two Hamiltonian terms are given by
\begin{subequations}
\begin{align}
\hat{H}_{\rm hop} &= -\frac{J}{2}\sum_{l=1}^{L-1} 
\hat{b}_{l+1}^\dagger\hat{b}_{l} + {\rm h.c.} \\ 
\hat{H}_{\rm int} &= \frac{U}{2}\sum_{l=1}^L\hat{n}_l(\hat{n}_l - 1),
\end{align}
\end{subequations}
where $\hat{b}_l$ ($\hat{b}_l^\dagger$) is an annihilation (creation) operator 
for bosons on site $l$, $\hat{n}_l = \hat{b}_l^\dagger\hat{b}_l$ is the bosonic 
number operator on site $l$, $L$ is the number of lattice sites and $N$ the 
number of bosonic particles on the lattice. We will consider BH systems with either unit filling, $N/L = 1$, or nearly unit filling. In this model the 
many-body quantum state lives in a Hilbert space of dimension 
\begin{equation}
 \label{eqn:BH_dim}
 \mathcal{D}_{\rm BH} = \frac{\left(N + L - 1\right)!}{N!(L-1)!},
\end{equation}
thus we will explore statistical complexity of diagonal ensembles, \textit{i.e.}, probability vectors, living in a $\mathcal{D}_{\rm BH}$-dimensional space. Equilibrium~\cite{Sengupta2005,Danshita2011} and 
nonequilibrium~\cite{Kollath2007,Sorg2014,Heyl2019BH} aspects of the dynamics of bosons in optical lattices have been extensively studied with this model. In particular, an experimental verification of the volume law characterizing thermal-like states for the Renyi entropy was presented in~\cite{Kaufman2016}, and an extensive study of equilibration and thermalization in the BH model was presented in~\cite{Sorg2014}. In contrast with the latter here we focus on the regime $J/U>1$ for which the BH model is known to be quantum chaotic~\cite{Kolovsky2004a,Cruz2020,Russomanno2020,Paush2022}.

\section{Results of numerical experiments}
\label{sec:preliminary_results}
In this section we present numerical results on the dynamics of 
statistical complexity under pure equilibration, and results on the robustness of this dynamics under the presence of an external perturbation.
\subsection{Dynamics of statistical complexity}
\begin{figure}[t!]
 \centering{\includegraphics[width=0.48\textwidth]{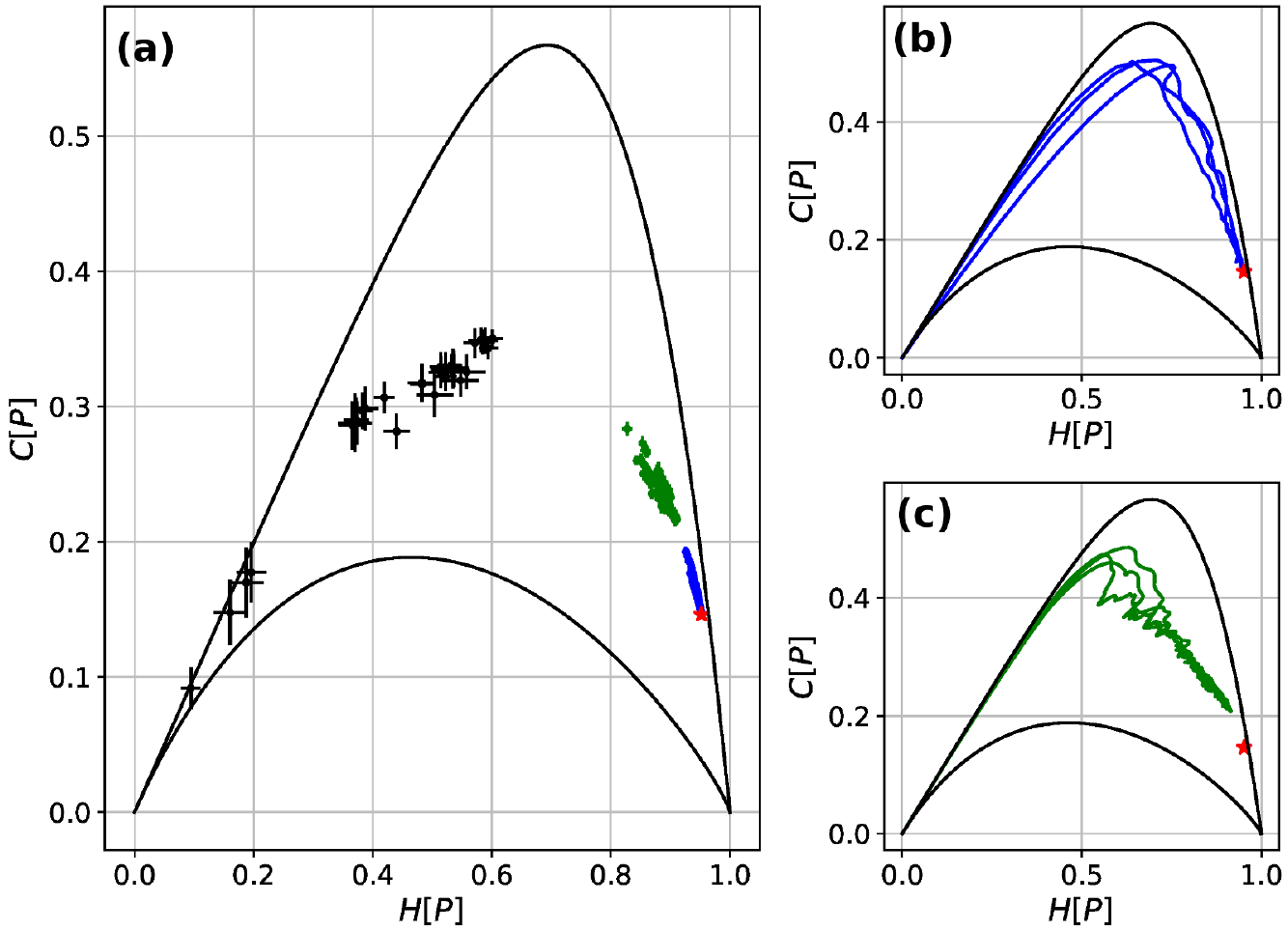}}
\caption{\textbf{(a)} Long time average E-C coordinates, $(\overline{H}, \overline{C})$, of the diagonal ensemble in the Fock basis, for all Fock states as initial states, quenched under $\hat{H}_{\rm BH}$ in Eq.~(\ref{eqn:BH_hamiltonian}). We work with a unit filled lattice at $N = L = 8$ and $U = 1.23$, $J = 2U$. \textbf{(b,c)} Dynamics of the diagonal ensemble in the Fock basis shown as trajectories in the entropy-complexity plane. Time flows from $H=0$ to $H=1$, following the monotonicity of the diagonal entropy. In both cases we show the trajectories of some representative initial Fock states, belonging to the states colored blue and green in (a), respectively. In all three subplots the red star shows the coordinates of distributions evolved under random unitaries sampled from the Circular Orthogonal Ensemble.}
\label{fig:E_C_plane_equilibrum}
\end{figure}
We study the dynamics of statistical complexity with the following setup. Initially one of the stationary states of $\hat{H}_{\rm int}$ is prepared, then time evolution is driven by $\hat{H}_{\rm BH}$ until a sufficient long time is reach and equilibrium is observed. In this section we will focus on a BH system with $8$ atoms in $8$ lattice sites and we fix $J = 2U$. Results for other system sizes are presented in App.~\ref{sec:app_other_results}.

Recall that the road to equilibrium is given by a hierarchy of diagonal ensembles, of the form
\begin{equation}
\label{eqn:road_equi_diago}
\rho|_{\rm diag}^{(t_0)}, \rho|_{\rm diag}^{(t_1)},..., \rho|_{\rm diag}^{(t_{\rm equiv})},...,\rho|_{\rm diag}^{(t_f)},
\end{equation}
with $t_f>t_{\rm equiv}$ some final time. In the E-C plane this hierarchy gives rise to a trajectory, defined by the succession of E-C coordinates 
\begin{widetext}
\begin{equation}
\label{eqn:E_C_trajectory}
(H[\rho|_{\rm diag}^{(t_0)}],C[\rho|_{\rm diag}^{(t_0)}]), (H[\rho|_{\rm diag}^{(t_1)}],C[\rho|_{\rm diag}^{(t_1)}]),..., (H[\rho|_{\rm diag}^{(t_{\rm equiv})}],C[\rho|_{\rm diag}^{(t_{\rm equiv})}]),..., (H[\rho|_{\rm diag}^{(t_f)}],C[\rho|_{\rm diag}^{(t_f)}]).
\end{equation}
\end{widetext}
Following the discussion in Sec.~\ref{sec:st_and_sc_plane} and using the generating functions in App.~\ref{sec:app_boundaries} we construct the boundary curves defining the E-C plane of our $8$in$8$ BH system. Then, For each initial Fock state we use their corresponding E-C trajectory to compute two complementary quantities. First the equilibrium E-C coordinate, obtained by approximating the long time average of both $H$ and $C$, defined as 
\begin{subequations}
\label{eqn:long_time_ec}
\begin{align}
\overline{H}[P] &= \lim_{T\to\infty} \frac{1}{T}\int_{0}^T H[P]dt,\\
\overline{C}[P] &= \lim_{T\to\infty} \frac{1}{T}\int_0^T C[P]dt.
\end{align}
\end{subequations}
These equilibrium coordinates are shown in the E-C plane as colorful dots in Fig.~\ref{fig:E_C_plane_equilibrum}a.

Using the E-C trajectory we define the criteria for equilibration as the size of the statistical variance of both $H$ and $C$ along the entire trajectory. If the state has equilibrated one expects these statistical fluctuations to be exponentially small in the system size. The size of the fluctuations can be seen as the error bars in Fig.~\ref{fig:E_C_plane_equilibrum}a. With this criteria we identify two sets of Fock states which do equilibrate (blue and green dots in Fig.~\ref{fig:E_C_plane_equilibrum}a) for which fluctuations are indeed exponentially small. These two sets of Fock states also correspond to states which are highly delocalized in the eigenbasis of the quench Hamiltonian. We also identified a set of states which do not equilibrate (blackdots in Fig.~\ref{fig:E_C_plane_equilibrum}), with large itinerant statistical fluctuations in their equilibrium E-C coordinates~\footnote{Since we are dealing with Fock states, these three different set of states can also be identified by the number of highly occupied single sites, however, we notice this is not a generic feature of the equilibration process we are studying, but rather a coincidence given the quantum system we selected as example.}.
We will go back to the discussion of localization/delocalization properties of equilibrating states later in this section.

\begin{figure}[t!]
 \centering{\includegraphics[width=0.485\textwidth]{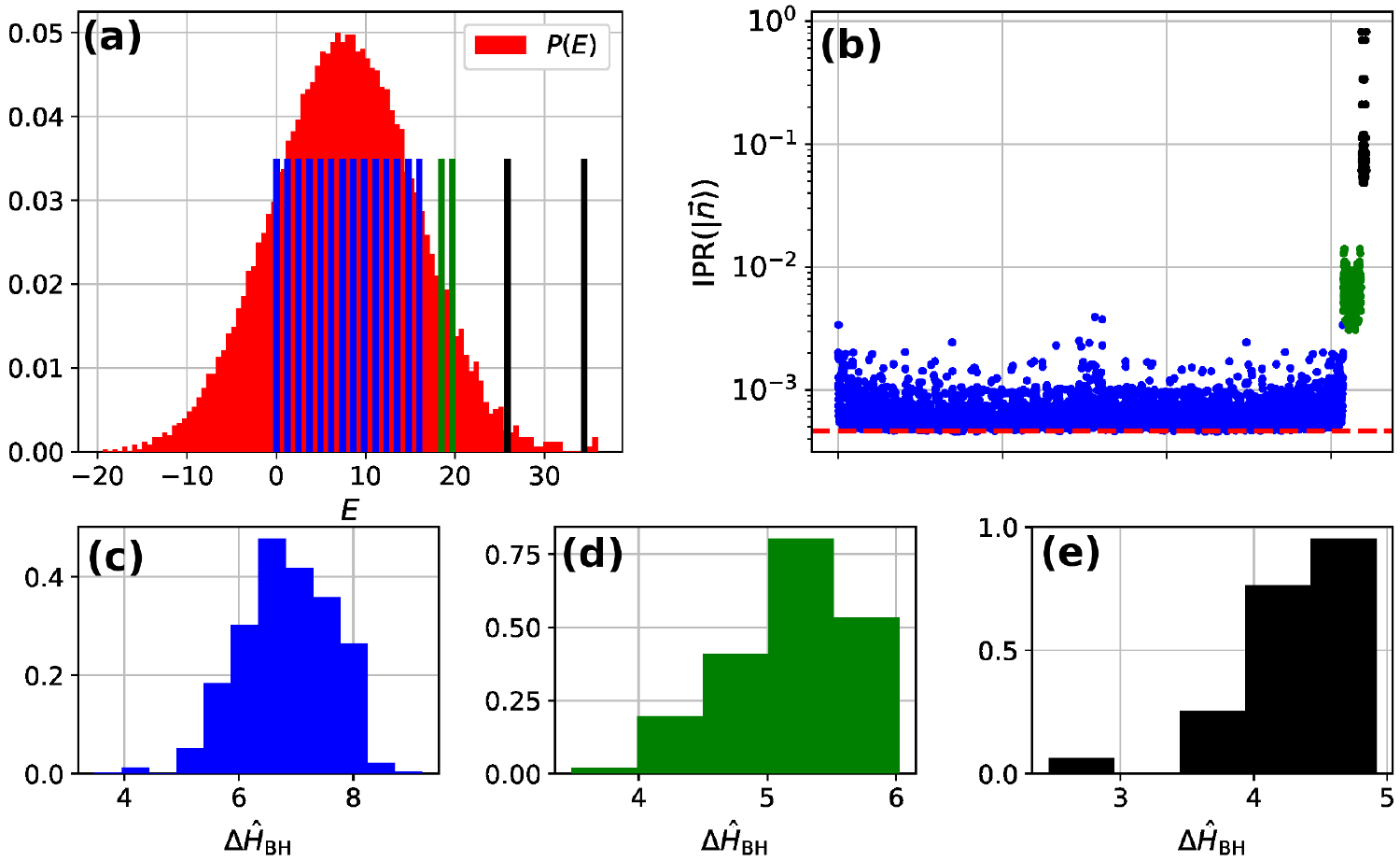}}
\caption{\textbf{(a)} Energy density of the quench BH Hamiltonian with $8$ atoms in $8$ lattice sites, and $J = 2U$ and $U = 1.23$. The energy density has a Gaussian shape, characteristic of quantum chaotic systems~\cite{Santos2011}. The bars represent the values of mean energy for all the states in the Fock basis. \textbf{(b)} Inverse participation ratio of the Fock states for the $8$ in $8$ BH system under study. The IPR is measured on the eigenbasis of $\hat{H}_{\rm BH}$. The dashed red line shows the IPR value averaged over the COE. \textbf{(c-e)} Width in energy of all the Fock states whose energy is shown in \textbf{(a)}. In all the subplots we use the same color code as in Fig.~\ref{fig:E_C_plane_equilibrum}.}
\label{fig:energy_densities}
\end{figure}

From the two sets of Fock states which equilibrate, blue and green dots in Fig.~\ref{fig:E_C_plane_equilibrum}a, we can readily identify those states which correspond more closely to ergodic states. This is done by comparing their E-C coordinate with the E-C coordinate of a random vector obtained by evolution under a random unitary sampled from the Circular Orthogonal Ensemble (COE). Which is shown by the red star in Fig.~\ref{fig:E_C_plane_equilibrum}a. We then recognize the blue set of Fock states as those resembling more closely ergodic states, and having the random vector as a good representation of their equilibrium properties~\footnote{Notice that the correspondence with random vectors will only be perfect in the thermodynamic limit.}. Additionally, all the states in the green and blue sets exhibit a larger value of normalized diagonal entropy when compared to those states which do not equilibrate. This last observation follows from the typical behavior of ergodic states, which are usually delocalized~\cite{Neuenhahn2012}. To investigate the localization properties of states in the basis of the quench Hamiltonian we computed the inverse participation ratio (IPR), defined as
\begin{equation}
\label{eqn:ipr}
{\rm IPR}(|\vec{n}\rangle) = 
\sum_{j=1}^{\mathcal{D}}|\langle\vec{n}|E\rangle|^4, 
\end{equation}
where $\vec{n} = (n_1, n_2,...,n_L)$ is a vector giving the occupations of the $L$ lattice sites for a given Fock state, with $\sum_{l=1}^{L}n_l = N$, and $\{|E\rangle \}$ is the eigenbasis of the quench Hamiltonian $\hat{H}_{\rm BH}$. The IPR in Eq.~(\ref{eqn:ipr}) gives $1$ for a fully localized state and $\frac{3}{\mathcal{D}_{\rm BH}+1}$ for a fully delocalized state. The latter correspond to the IPR value averaged over states obtained from the eigenbasis of random Hamiltonians sampled from the Gaussian Orthogonal Ensemble (GOE) (see the methods in~\cite{Sieberer2019} for further details). Results for the $8$ in $8$ BH system are shown in Fig.~\ref{fig:energy_densities}b, we have ordered the Fock basis such that states in each of the three sets, blue, green and black, appear together. All the states in the blue set are saturating the COE value, implying maximal delocalization and hence and expected ergodic behavior~\cite{Neuenhahn2012}, states in the black set show IPR's which are two orders of magnitude larger than those in the blue set, thus we expect strong nonergodic behavior with long lived coherences. Finally, states in the green set are characterized by IPR values in between those of the blue and black sets, however these values cluster near those of the blue set, indicating a stronger presence of ergodic properties, and as such we see them reaching equilibration.

In Fig.~\ref{fig:E_C_plane_equilibrum}b,c we display the E-C 
trajectories followed by a subset of states in each of the two sets reaching equilibrium, \textit{i.e.}, blue and green states in Fig.~\ref{fig:E_C_plane_equilibrum}a. Few common universal features can be appreciated. First, for small values of the normalized entropy, \textit{i.e.}, short times, the diagonal ensembles closely resemble $P_{\rm upper}$, thus have forms similar to those in Eq.~(\ref{eqn:boundary_2}). This monotonic character of the statistical complexity reaches a maximum for intermediate values of the normalized diagonal entropy, after which trajectories turn around and asymptotically tend towards the E-C coordinate of the COE. Therefore, equilibration dynamics characterized to be an ever increasing entropy process generates complex structures only for a portion of the local equilibration time, after which those structures are washed out as the distributions start resembling a random one. 

Similarly two differences between Fock states in the blue and green sets are manifested. First, those in the blue set peaked at higher values of statistical complexity and equilibrate to the COE value, whereas those in the green set have a smaller maximum of statistical complexity and do not saturate the COE value.

In Fig.~\ref{fig:energy_densities}a we show the energy density of the quench Hamiltonian (red histogram), defined as
\begin{equation}
\label{eqn:quench_energy_density}
P(E) = \sum_{l=1}^{\mathcal{D}_{\rm BH}} \delta(E - E_l),
\end{equation}
where $\hat{H}_{\rm BH}|E_l\rangle = E_l|E_l\rangle$. We notice the Gaussian shape of the energy density, characteristic of quantum chaotic Hamiltonians~\cite{Santos2011}. On top of the energy density histogram in Fig.~\ref{fig:energy_densities}a we show the mean energies, $\langle\vec{n}|\hat{H}_{\rm BH}|\vec{n}\rangle$, of each and all the Fock states. The height and width of the blue, green and black bars are arbitrary, they only illustrate the location of the mean energies for the different Fock states. States in the blue set, which equilibrate close to the COE value, have mean energies deep within the bulk and denser part of the energy spectrum, or in other words, when expressed in the eigenbasis of the quench Hamiltonian will have support on a large number of eigenvectors, they are ``delocalized'' in the energy basis as follows from the IPR results in Fig.~\ref{fig:energy_densities}b. On the other hand, the states in the black subset, which do not equilibrate, are all very close to the edge of the spectrum, in a sense they are ``localized'' in energy, hence coherences are long lived. Finally, the states on the green subset have mean energies in between those of the blue set and those of the black set. Thus, they are neither fully ``localized'' nor completely ``delocalized'' in energy, and display a mixture of properties between those of ergodic and nonergodic states. 

The previous analysis is complemented with the energy width of all the Fock states, that is, $\Delta \hat{H} = \sqrt{\langle \vec{n}| \hat{H}_{\rm BH}^2|\vec{n} \rangle - \langle \vec{n}| \hat{H}_{\rm BH}|\vec{n} \rangle^2}$. Results are shown in the histograms of Fig.~\ref{fig:energy_densities}c-e, from this results is evident that the blue states are the ones with wider support on the energy basis of all, and thus diffusion in Hilbert space is facilitated. Those in the black set are the narrower ones. Notice how,  both, the maximum and mean of the histograms moves to smaller values of $\Delta \hat{H}$ blue in Fig.~\ref{fig:energy_densities}c to black in Fig.~\ref{fig:energy_densities}e. Similar to the IPR results in Fig.~\ref{fig:energy_densities}b, the width of the green states is closer to those of the blue states than those of the black states. Hence, they display properties which are a mix between ergodic and nonergodic  behaviors, yet the ergodic behavior is dominant and thus we see these states equilibrating.

\subsection{The effect of a random perturbation}
In the previous section we discussed the road E-C trajectories of all the initial Fock states, and we used these trajectories to characterize the road to equilibrium. We found a subset of the initial Fock states which are ergodic states. This implies that the form of the E-C trajectories in Fig.~\ref{fig:E_C_plane_equilibrum}b is a generic feature of the road to equilibrium for any typical (ergodic) state. We also observe the E-C coordinates of these ergodic states to be fairly close to the E-C coordinate of the COE. This is another generic feature of ergodic states, as their equilibrium properties are usually well described by the appropriate random matrix ensemble. Follwoing from these results, in this section we will explore the question: how the road to equilibrium generated by a physical chaotic Hamiltonian differs from the one driven by a random Hamiltonian? We study this question in two different ways. First, by looking at how the presence of a perturbation in the form of a random Hamiltonian modifies the E-C trajectories shown in Fig.~\ref{fig:E_C_plane_equilibrum}b,c and the difference of these Hamiltonian E-C trajectories with those coming from the evolution generated by a random Hamiltonian. In particular we will show evidence of a separation between nonequilibrium Hamiltonian evolution and random evolution. The former produces E-C trajectories composed of diagonal ensembles with higher values of statistical complexity. Second, the robustness of the observed E-C plane trajectories to the action of the random perturbation as a function of the perturbation strength.

\subsubsection{Effects of a random perturbation in the Fock basis}
\begin{figure}[t!]
 \centering{\includegraphics[width=0.485\textwidth]{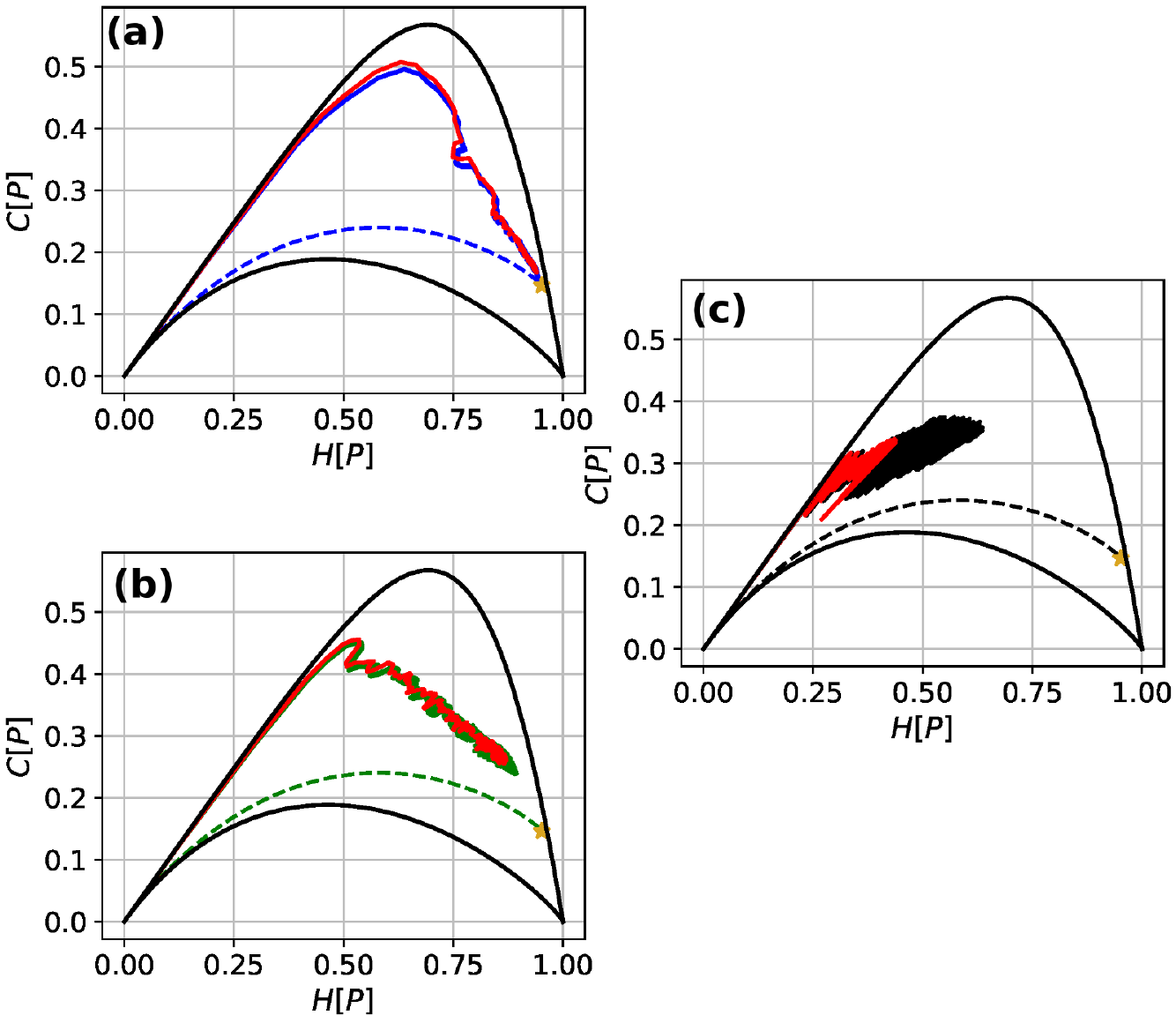}}
\caption{Trajectories in the E-C plane of the diagonal ensemble in the Fock basis, for a single initial Fock state evolved under the Hamiltonian in Eq.~(\ref{eqn:perturbed_BH_hamiltonian}), for each of the three sets identified in Fig.~\ref{fig:E_C_plane_equilibrum}a, blue \textbf{(a)}, green \textbf{(b)}, and black \textbf{(c)}. In all panels, the continuous red line corresponds to the unperturbed E-C trajectory, the dashed line to the E-C trajectory obtained evolving under a random Hamiltonian, and the yellow star the COE coordinate. The continuous, blue, green and black lines show $40$ different trajectories resulting from $40$ different realization of the perturbed evolution, each with a different random Hamiltonian. For the displayed results we use $\epsilon = 0.5$ for the perturbation strength, other parameters are as in Fig.~\ref{fig:E_C_plane_equilibrum}.}
\label{fig:pertu_number_basis}
\end{figure}
Let us consider a perturbation, in the form of a random Hamiltonian, 
acting on the quench Hamiltonian. Then equilibration dynamics is driven by a Hamiltonian of the form
\begin{equation}
 \label{eqn:perturbed_BH_hamiltonian}
 \hat{H}_{\rm BH}' = \hat{H}_{\rm BH} + \epsilon\Delta_0 \hat{H},
\end{equation}
where $\hat{H}_{\rm BH}$ is defined in Eq. (\ref{eqn:BH_hamiltonian}), 
$\hat{H}$ is a random Hamiltonian sampled from the GOE, $\epsilon$ a 
parameter controlling the strength of the perturbation, and $\Delta_0$ the mean 
energy level spacing of $\hat{H}_{\rm BH}$.

After adding the random Hamiltonian, we evolve different initial Fock states 
belonging to each of the three different sets identified in Fig.~\ref{fig:E_C_plane_equilibrum}a, and construct the 
perturbed trajectories in the E-C plane. They are shown in Fig~\ref{fig:pertu_number_basis}a-c, where the trajectories for $40$ different 
random Hamiltonians are shown for a single initial Fock state, along with the 
original trajectory (continuous red lines).

The dashed lines in Fig.~\ref{fig:pertu_number_basis} correspond to the evolution in the E-C plane of initial Fock states under a random Hamiltonian of the form $\epsilon\Delta_0\hat{H}$. We compare this trajectory to the one obtained under the Hamiltonian evolution, and immediately recognize a clear separation of the two roads to equilibrium given by the E-C trajectories. Hamiltonian evolution builds a road towards equilibration constituted by diagonal ensembles characterized by large values of statistical complexity, remaining close to the upper boundary of the E-C plane for small values of the entropy. On the other hand, the evolution dominated by a random Hamiltonian builds a road to equilibrium via a succession of diagonal ensembles which lack structure, and thus the E-C trajectory remains, for all values of the entropy, close to the lower boundary curve of the E-C plane. Furthermore, evolution under the perturbed Hamiltonian in Eq.~\ref{eqn:perturbed_BH_hamiltonian} generates a E-C trajectory which interpolates between that of a pure Hamiltonian evolution and that of an evolution under a random Hamiltonian, as a function of the perturbation strength $\epsilon$.

The E-C trajectories describing the road to equilibrium are a low dimensional representation of the evolution towards equilibrium of the diagonal ensembles, which are probability vectors in a $\mathcal{D}_{\rm BH}$-dimensional probability space. As such, the difference between the road to equilibrium under coherent quantum evolution and under evolution by a random Hamiltonian can be further investigated in the behavior of their respective trajectories in probability space.

The space of $\mathcal{D}$-dimensional probability vectors is the set of all the convex combinations of the linearly independent vectors $\{\vec{e}_j\}_{j=1,..,\mathcal{D}}$, given by
\begin{equation}
 \label{eqn:prob_simplex}
 \mathbb{P}^{\mathcal{D}-1} = \left\{ 
\vec{v}\in\mathbb{R}^{\mathcal{D}}:\vec{v}= \sum_{j=1}^{\mathcal{D}} 
u_j\vec{e}_j,\enspace \sum_{j=1}^{\mathcal{D}} u_j=1,\enspace u_j\ge0 \right\}.
\end{equation}
A set of real vectors satisfying Eq.~(\ref{eqn:prob_simplex}) is called a $(\mathcal{D}-1)$-simplex. 

The vertices of this $(\mathcal{D}-1)$-simplex are the basis vectors $\vec{e}_l$, and are represented in the E-C plane at the $(0,0)$ point (left corner), having multiplicity $\mathcal{D}$. The ``infinite temperature'' state can be identified with the baricenter of $\mathbb{P}^{\mathcal{D}-1}$, it is unique and is represented by the $(1,0)$ point at the right corner on the E-C plane. It can be shown that the behavior of $C$ is identical in each one of the $\mathcal{D}!$ baricentric subdivisions of $\mathbb{P}^{\mathcal{D}-1}$, see Ref.~\cite{Martin2006}. A baricentric subdivision of $\mathbb{P}^{\mathcal{D}-1}$ is the subset of probability vectors
\begin{equation}
\label{eqn:baricenter_sub}
\mathbb{P}^{\mathcal{D}-1}_{\rm bari} = \left\{ \vec{g}_{\tau}^{(m)}: 
\tau\in{\rm Permu}(\mathcal{D}),\enspace m=1,...,\mathcal{D} \right\}, 
\end{equation}
where Permu$(\mathcal{D})$ is the set of permutations of $[1,2,..,\mathcal{D}]$ 
and the vectors $\vec{g}_{\tau}^{(m)}$ are given by 
\begin{equation}
\vec{g}_{\tau}^{(m)} = \sum_{n=\tau(m)}^{\tau(N)} 
\left(\frac{1}{\mathcal{D}-m+1}\right)\vec{e}_{\tau(n)}, 
\end{equation}
with ${\vec{e}_j}$ the vertices of $\mathbb{P}^{\mathcal{D}-1}$. In our numerical experiments, for a given Fock state represented by $\vec{e}_l$, we are choosing the baricentric subdivision of $\mathbb{P}^{\mathcal{D}-1}$ such that $\tau$ is a permutation swapping $\mathcal{D}$ with $l$, therefore the subsimplex $\mathbb{P}^{\mathcal{D}-1}_{\rm bari}$ has $\rho_\infty$ as the first vertex and $\vec{e}_l$ as the last one.

The distributions giving the two boundary curves of the E-C plane correspond to the probability vectors on the edges of the baricentric subdivision of $\mathbb{P}^{\mathcal{D}-1}$ for the Fock state under study. In particular, for our choice of entropy and disequilibrium, distributions on the edge connecting the very first and last (the two nonconsequetive) vertices build the lower boundary curve of the E-C plane, see Eq.~(\ref{eqn:boundary_1}). Those distributions on the union of all the edges connecting consecutive vertices build the upper boundary curve of the E-C plane, see Eq.~(\ref{eqn:boundary_2}). We saw in Fig.~\ref{fig:E_C_plane_equilibrum}a that the equilibrium state is obtained by the action of a COE unitary on one of the basis vectors $\vec{e}_j$, for a given baricentric subdivision, this equilibrium state is a distribution in the vicinity of the baricenter, \textit{i.e.}, the ``infinite temperature'' state. Thus, in probability space reaching equilibrium means the trajectory is already inside this neighborhood of the baricenter and remains there for all later times. 

We can use these correspondences to  create a picture in the $(\mathcal{D}-1)$-simplex of the trajectories for each of the two time evolutions under scrutiny. Consider first the evolution driven by a random Hamiltonian. We saw in Fig.~\ref{fig:pertu_number_basis}a-c that the E-C trajectory remains for all values of $H[\rho|_{\rm diag}^{(t)}]$ close to the lower bound curve. Therefore in probability space, the initial state $\vec{e}_j$ moves towards the neighborhood of the baricenter staying at all times close to the edge connecting the baricenter with $\vec{e}_j$ until reaching equilibrium. On the other hand the quantum evolution remains close to the upper bound curve on the E-C plane (see Fig.~\ref{fig:E_C_plane_equilibrum}b,c), thus the initial state $\vec{e}_j$ evolves along a path that follows all the edges connecting consecutive vertices on the baricentric subdivision, until reaching the neighborhood of the baricenter and equilibrating.    

\subsubsection{Robustness analysis in the basis of the quench Hamiltonian}
\begin{figure}[t!]
\centering{\includegraphics[width=0.485\textwidth]{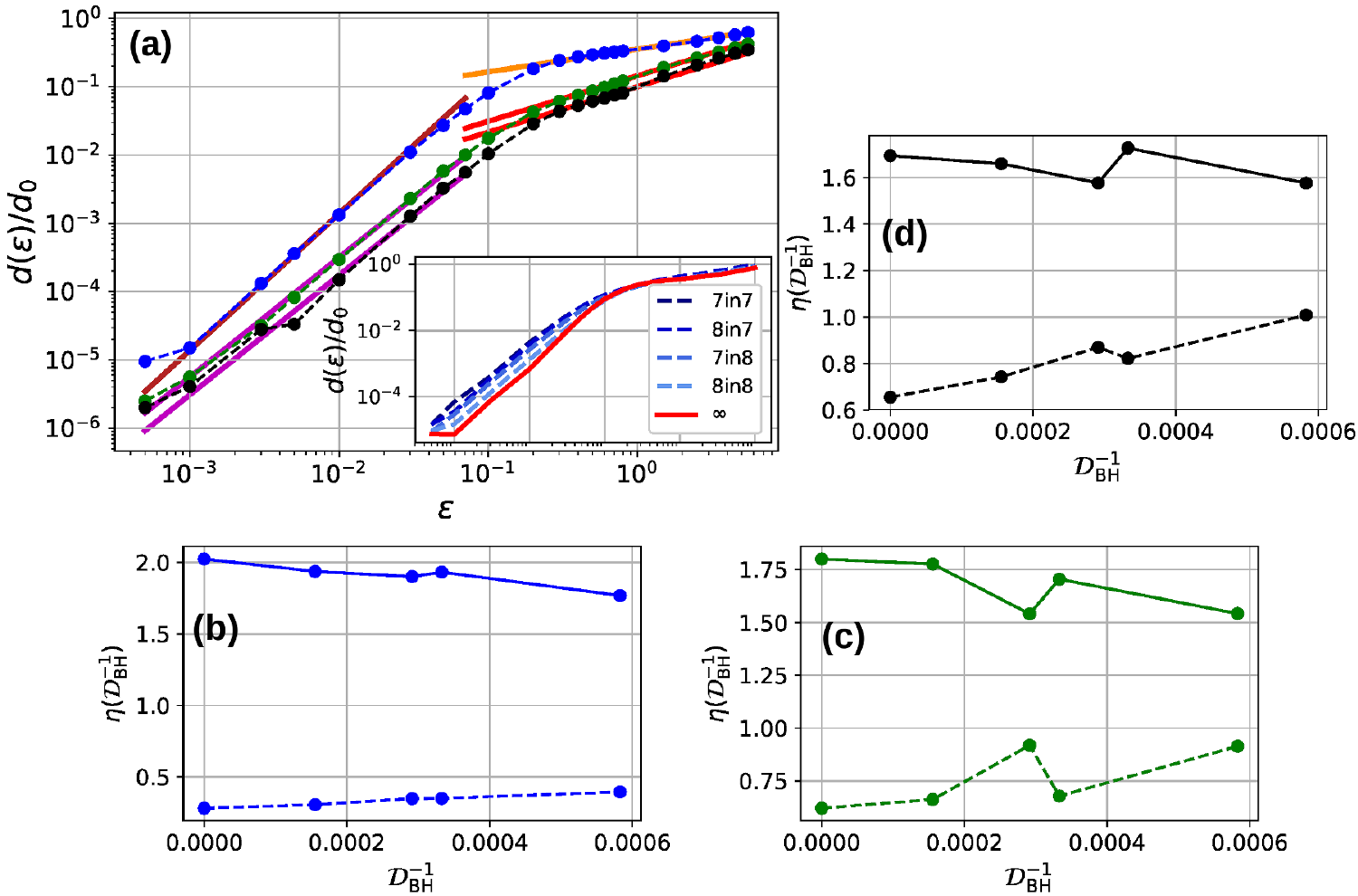}}
\caption{\textbf{(a)} Euclidean distance between the diagonal ensemble of Fock states, in the basis of the quench Hamiltonian, to their diagonal ensemble after the action of a random unitary. The normalization factor is given by the distance between the E-C coordinate of $\rho|^{(t_0)}_{\rm diag} = |c_l(0)|^2$ and the E-C coordinate of the COE. Each point is the result of averaging over ten initial Fock states in each of the three sets (blue, green and black). The distance to the COE coordinate for each of those states is the result of averaging over $100$ different trajectories generated by $100$ random Hamiltonians sampled from the GOE.} The inset shows the $d(\epsilon)$ for all the BH systems considered, the solid red line is the finite size scaling curve giving the infinite size limit. The inset and the main subplot share the $x$-axis. \textbf{(b-d)} Characteristic exponent $\eta$ of the sensitivity to perturbations as a function of the inverse dimension $1/\mathcal{D}_{\rm BH}$. In all figures the solid line represents the points in the regime $\epsilon<0.1$ and the dashed lines the points in the regime $\epsilon>0.1$. The point at $1/\mathcal{D}_{\rm BH}=0$ is obtained from the finite size scaling.
\label{fig:distance_energy_basis}
\end{figure}
In order to gauge the robustness of our set of initial states against a random perturbation, we need to nullify the effect of the nonequilibrium dynamics. This can be readily done by choosing the eigenbasis of the quench Hamiltonian, $\hat{H}_{\rm BH}$ as the reference basis. In this basis, an initial Fock state can be written as 
\begin{equation}
 \label{eqn:fock_in_energy_basis}
 |n_1n_2...n_L\rangle = \sum_{l=1}^{\mathcal{D}_{\rm BH}} c_l(0) |E_l\rangle,
\end{equation}
where $c_l(0) = \langle n_1n_2...n_L|E_l\rangle$. Here, the initial diagonal ensemble $\rho|_{\rm diag}^{(t_0)} = |c_l(0)|^2$ is invariant to the action of the unitary $e^{-it\hat{H}_{\rm BH}}$, and the E-C trajectories are now represented by a single point. Under the action of a random unitary of the form $\hat{U}_{\rm rand}(\epsilon) = e^{-it\epsilon\Delta_0\hat{H}}$ with $\hat{H}$ a random Hamiltonian as in Eq.~\ref{eqn:perturbed_BH_hamiltonian}, the diagonal ensemble does evolve. Then, as a function of the perturbation strength, $\epsilon$, the diagonal ensemble traces a new E-C trajectory going from $\rho|_{\rm diag}^{(t_0)} = |c_l(0)|^2$ to the E-C coordinate of the COE. We see then that quantifying the robustness of a given initial Fock state can be done via the euclidean distance between the E-C coordinate of $\rho|_{\rm diag}^{(t_0)} = |c_l(0)|^2$, $\left(H[|c_l(0)|^2], C[|c_l(0)|^2]\right)$, and the E-C coordinate of the diagonal ensemble after the action of the random Hamiltonian, $\rho|_{\rm diag}^{\rm rand} = |c_l({\rm rand})|^2$, given by $\left(\overline{H}[|c_l({\rm rand})|^2], \overline{C}[|c_l({\rm rand})|^2]\right)$, where $c_l({\rm rand}) = \langle n_1n_2...n_L|\hat{U}_{\rm rand}(\epsilon)|E_l\rangle$ and the overline indicates average over random unitaries. In other words, we look at the distance
\begin{equation}
 \label{eqn:distance_energy_basis}
 d(\epsilon) = \sqrt{\Delta H^2(\epsilon) + \Delta C^2(\epsilon)},
\end{equation} 
where $\Delta H = \overline{H}[|c_l({\rm rand})|^2] - H[|c_l(0)|^2]$ and $\Delta C = \overline{C}[|c_l({\rm rand})|^2] - C[|c_l(0)|^2]$. In Fig.~\ref{fig:distance_energy_basis} we show the normalized distance to the COE coordinate as a function of the perturbation strength $\epsilon$ for each of the three sets of states identified in Fig.~\ref{fig:E_C_plane_equilibrum}a. For each set of states, we compute this distance as the result of two averages, one over a subset of $10$ states and one over $100$ different random Hamiltonians. The normalization factor $d_0$ corresponds to the euclidean distance between the E-C coordinate of $\rho|_{\rm diag}^{(t_0)} = |c_l(0)|^2$ and the E-C coordinate of the COE.

We notice that for a large enough $\epsilon$, all initial Fock states are mapped to a random state by the action of $\hat{U}_{\rm rand}(\epsilon)$, as can be observe by the distance tending to one in the right most end of Fig.~\ref{fig:distance_energy_basis}a. As the the value of $\epsilon$ is decrease we observe two different behaviors of the distance to the E-C coordinate of the initial diagonal ensemble, one for small $\epsilon<0.1$ and one for large $\epsilon>0.1$. The linear behavior of the distance in each of these two regimes, on a log-log scale, suggests that the sensitivity to a perturbation in the form of a random Hamiltonian exhibits a characteristic exponent. In other words, one has,  $\frac{d(\epsilon)}{d_0}\sim\epsilon^\eta$ with $\eta$ the characteristic exponent. The typical states (blue dots in Fig.~\ref{fig:E_C_plane_equilibrum}a) present a higher sensitivity to the effect of the random perturbation in the regime $\epsilon<0.1$, where the growth rate in this regime goes as $\eta\approx2$ (see dark red line in Fig.~\ref{fig:distance_energy_basis}a), indicating that typical states (ergodic states) are more fragile to the effects of an added random perturbation. As a consequence of this larger sensitivity in the regime $\epsilon<0.1$, the set of blue states exhibits the slower growth rate in the regime $\epsilon>0.1$, given by $\eta\approx1/3$, since saturation to the COE coordinate happens faster. In contrast the green and black states exhibit a smaller sensitivity to the effects of the random perturbation in the regime $\epsilon<0.1$, where we obtained the dependence $\eta\approx7/4$, and as a consequence exhibit a larger sensitivity in the regime $\epsilon>0.1$, in the process of saturating to the COE coordinate. For this second regime we obtained the dependence $\eta\approx2/3$ (see the magenta and red lines in Fig.~\ref{fig:distance_energy_basis}a).

In order to complete the characterization of sensitivity to random perturbations using the E-C plane, we obtained numerical results for other three B-H systems with filling factors $N/L = 7/7, 8/7, 7/8$, respectively. The equilibrium E-C coordinates and IPR values for these other B-H systems are shown in App.~\ref{sec:app_other_results}. The respective growth rates were obtained by similar means as those in Fig.~\ref{fig:distance_energy_basis}a. We use these four data points, as a function of the inverse dimension $1/\mathcal{D}_{\rm BH}$ to find the value of $\eta$ in the thermodynamic limit $\mathcal{D}_{\rm BH}\to\infty$ via a finite size scaling. In other words, the intercept of a linear fit to the points of the form $(1/\mathcal{D}_{\rm BH}, \eta)$. Results are shown in Fig.~\ref{fig:distance_energy_basis}b-d, where we have included the result of the finite size scaling, point at $1/\mathcal{D}_{\rm BH}=0$. These results confirm our previous observation for the states in the blue set, with the two sensitivity regimes $\eta=2$ and $\eta=1/3$ for $\epsilon<0.1$ and $\epsilon>0.1$, respectively. On the other hand the results of the finite size scaling point towards a small separation in sensitivity between green and black states in the regime $\epsilon<0.1$, we found $\eta=9/5$ and $\eta=17/10$ for green and black, respectively. We adjudicate this to the fact that the properties of the states in the green set are a mixture between ergodic and nonergodic. Whereas in the regime $\epsilon>0.1$ we obtained $\eta=2/3$ for both sets, supporting our previous findings.

\section{Summary and outlook}
\label{sec:final_remarks}
In this work we have revisited the paradigmatic problem of an interacting many-body quantum system equilibrating under its own dynamics. We focus on the situation of equilibration as a consequence of a quantum quench, and study the time evolution of diagonal ensembles. We showed, using tools from information theory and the theory of majorization, that the equilibration process in quantum chaotic many-body systems can be understood as the emergence of a hierarchy of diagonal ensembles, inducing a hierarchization of probability space. We referred to this hierarchy of diagonal ensembles as the road to equilibrium, and investigated its structure in detail. We showed that this hierarchization guarantees the emergence of a form of the second law of thermodynamics for the diagonal entropy. This latter fact allowed us to use some statistical tools, namely the statistical complexity and the entropy-complexity plane, popular in the analysis of time series data, to investigate salient features of the road to equilibrium.

As an illustrative example we focus on a system of bosons trapped in a one dimensional lattice, described by the single band Bose-Hubbard model. The analysis of the road to equilibrium as an E-C trajectory revealed that initial states with ergodic properties visit diagonal ensembles of high statistical complexity during their road to equilibrium. Furthermore, these trajectories all owe us to present a clear separation between the road to equilibrium driven by a physical quantum Hamiltonian and that generated by evolution under a random Hamiltonian, which nevertheless captures almost perfectly the system properties once equilibrium has been reached. In the road towards equilibrium the former explores a region of probability space characterized by a large value of statistical complexity, whereas the latter visits states with small values of statistical complexity, staying at all times very close to the allowed lower bound. Finally, we use the E-C plane in the basis of the quench Hamiltonian to characterize the robustness of the quantum dynamics against the effects of the random perturbation, finding that all the initial Fock states, either ergodic or not, exhibit a polynomial sensitivity.

The discussion in this manuscript provides a new point of view into well established results in the study of many-body quantum systems out-of-equilibrium. Additionally, they illustrate the usefulness of certain statistical tools, popular in the study of time series and classical dynamical systems, in the study and characterization of out-of-equilibrium many-body quantum dynamics, in particular, opening the door to a convergence of quantum many-body dynamics and complex systems science.

In the present manuscript we focused on the study of diagonal ensembles and ignored the off-diagonal elements of the density matrix. However, the off-diagonal elements carry important information of the nonequilibrium dynamics. In principle, by constructing a well defined probability distribution based on this elements one could use the tools presented in this manuscript to characterize their behavior during a nonequilibrium process. This immediately poises the question of how can we extend the current analysis to use the information encoded in the full quantum state. An alternative, which is currently under investigation, is to consider reduced density matrices by partially tracing out part of the lattice system. In particular, when considering reduced states of many-body lattice systems, one could exploit the entropy-complexity plane to distinguish scrambling dynamics from pure decoherence, a topic which has seen elevated interest in recent years~\cite{Yoshida2019,Touil2021}.

\acknowledgements
The author extends his gratitude to Anupam Mitra, for his enthusiasm towards this work, several useful discussions and his careful reading of the manuscript. The author is also grateful to Carlos A. Parra-Murillo and Pablo M. Poggi for comments and discussions during an early stage of this project. I acknowledge the Boulder School for Condensed Matter and Material Physics 2021: ultracold matter, during which some final aspects of this work were completed. Most of the results presented in this manuscript were obtained between spring 2018 and summer 2019, period of time during which the author was financially supported through teaching assistantships from the Physics and Astronomy Department at the University of New Mexico.

\appendix
\section{Boundary curves of the entropy-complexity plane}
\label{sec:app_boundaries}
The explicit form of the distributions defining the upper and lower bound curves of the entropy-complexity plane can be constructed from the following generating functions
\begin{equation}
 \label{eqn:boundary_1}
 \mathcal{P}_{1,\mathcal{D}}(r) = r\sum_{j=1}^{\mathcal{D}-1}\vec{e}_j - 
+\left(1 - r(\mathcal{D}-1)\right)\vec{e}_\mathcal{D},
\end{equation}
where $r\in\left[0,\frac{1}{\mathcal{D}}\right]$, and 
\begin{equation}
 \label{eqn:boundary_2}
 \mathcal{P}_{j,j+1}(r) = r\vec{e}_j + \left(\frac{1 - 
r}{\mathcal{D}-j}\right)\sum_{l=j+1}^\mathcal{D}\vec{e}_l
\end{equation}
where $r\in\left[0,\frac{1}{\mathcal{D}-j+1}\right]$ and $j=1,..,\mathcal{D}-1$. Here $\{\vec{e}_j\}_{j=1,..,\mathcal{D}}$ is the canonical basis of $\mathbb{R}^\mathcal{D}$, which constitutes a set of basis vectors for the space of $\mathcal{D}$-dimensional probability vectors under the $1$-norm. Depending on the details of the construction of $C$ one can have Eq.~(\ref{eqn:boundary_1}) generating distributions which saturate either of the two bounds, hence Eq.~(\ref{eqn:boundary_2}) will generate those saturating the other one. Regularity conditions allowing the identification of the boundary upper and lower distributions were laid out in~\cite{Martin2006}. For our choice of Jensen-Shannon divergence and Shanon entropy, distributions generated by Eq.~(\ref{eqn:boundary_1}) build $C[P_{\rm lower}]$ and 
distributions generated by Eq.~(\ref{eqn:boundary_2}) build $C[P_{\rm upper}]$.

\begin{figure}[ht!]
 \centering{\includegraphics[width=0.485\textwidth]{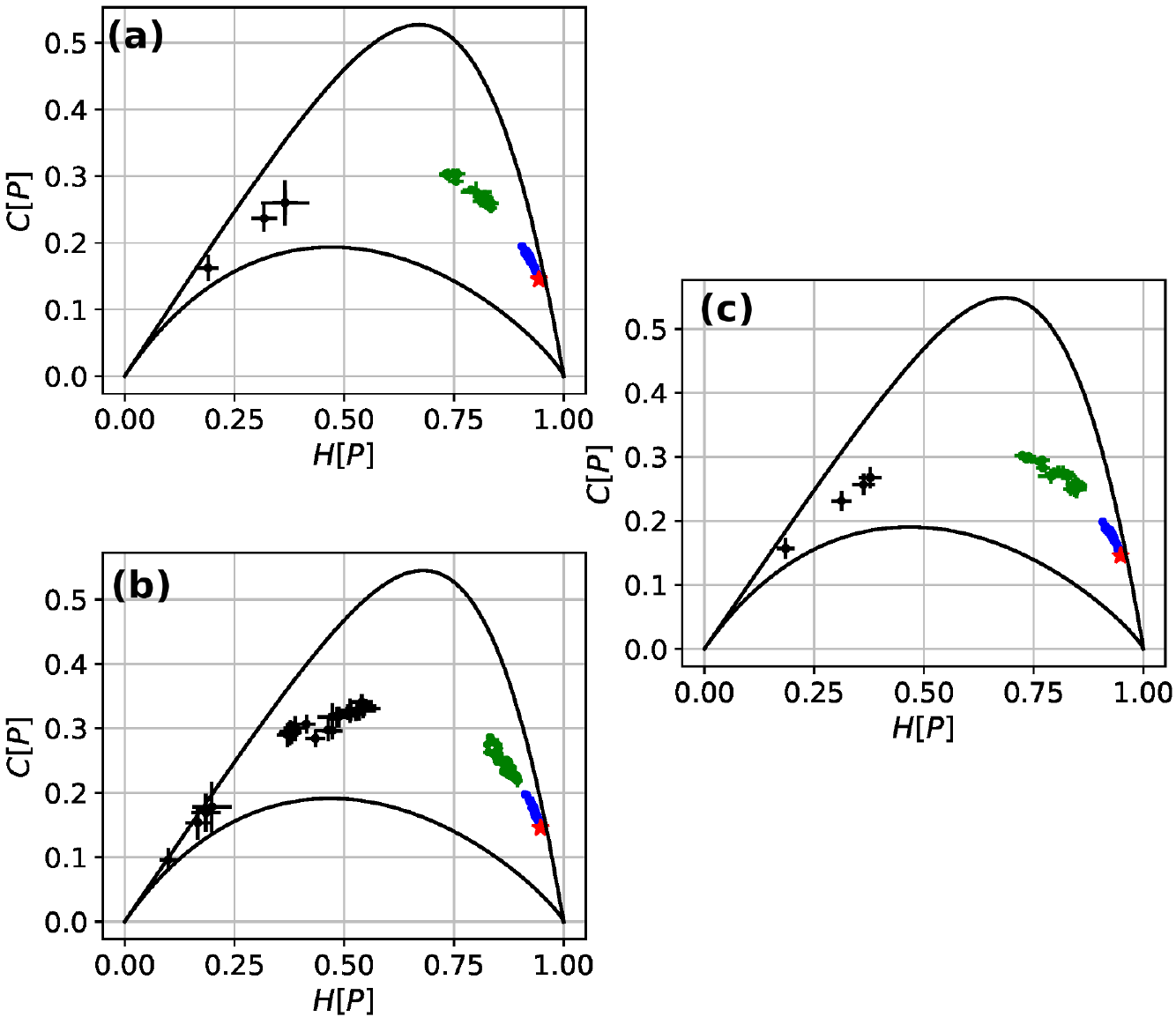}}
\caption{Equilibrium coordinates in the entropy-complexity plane for all the 
states in the Fock basis, after evolving under a Hamiltonian quench. The shown 
coordinates in \textbf{(a)}, \textbf{(b)}, \textbf{(c)} correspond to B-H 
systems with filling factors of $N/L = 7/7, 8/7, 7/8$, respectively. For all 
the studied BH systems we observe the separation, in the E-C plane, of the Fock 
basis into the three sets identified in Fig. \ref{fig:E_C_plane_equilibrum}a.}
\label{fig:more_planes}
\end{figure}
\begin{figure}[ht!]
\centering{\includegraphics[width=0.485\textwidth]{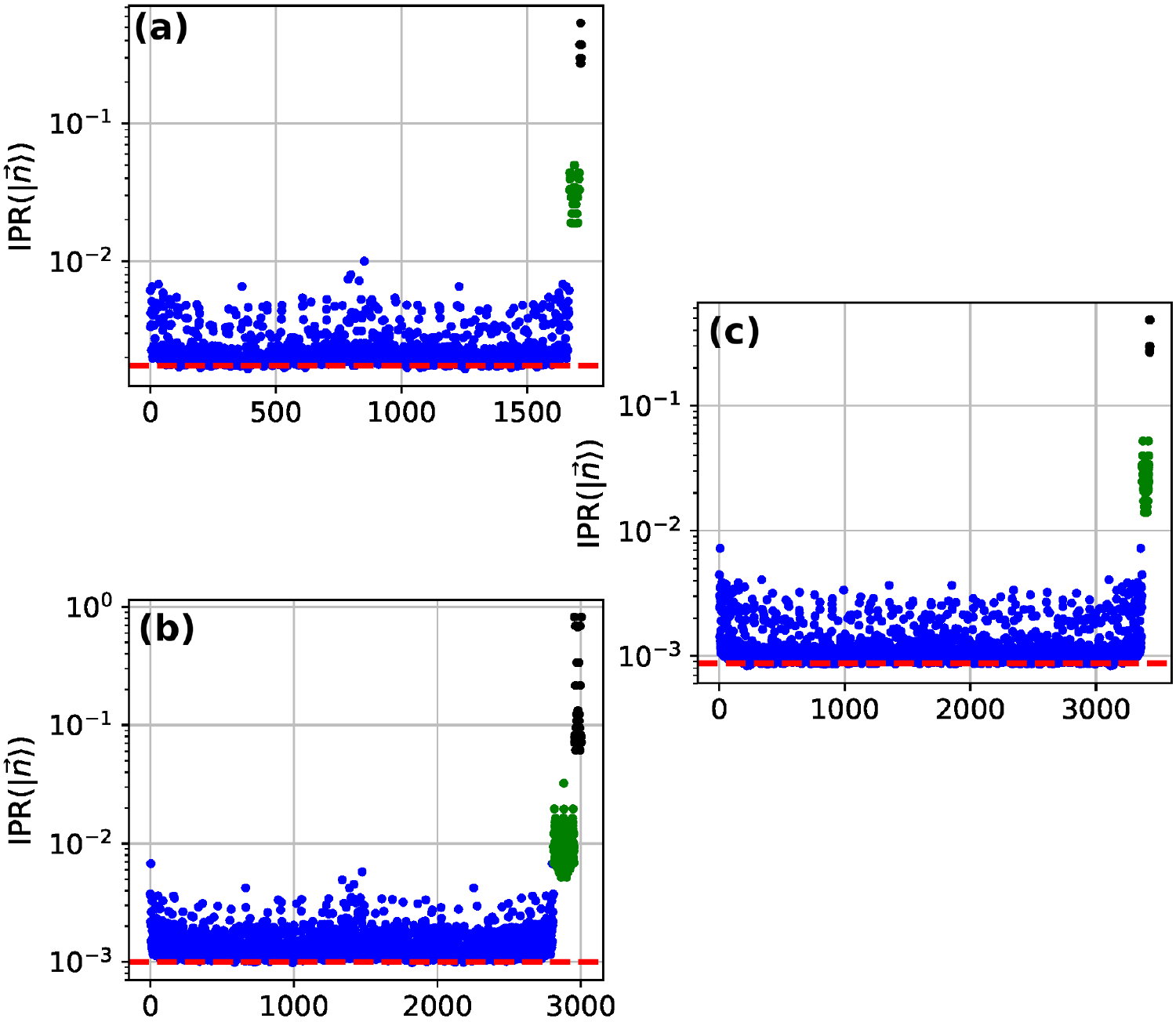}}
\caption{Inverse participation ratios of Fock basis states, measured in the basis of the quench Hamiltonian, $\hat{H}_{\rm BH}$. Results are for the BH systems with $N/L = 7/7, 8/7, 7/8$ in \textbf{(a)}, \textbf{(b)}, \textbf{(c)}, respectively. We have ordered the Fock basis so that states in each of the three sets, blue, green and black will appear together. The dashed red line shows the IPR value average over the COE.}
\label{fig:more_ipr}
\end{figure}

\section{Results for other system sizes}
\label{sec:app_other_results}
In this appendix we show additional results for the other BH systems considered in our study. In Fig.~\ref{fig:more_planes}a-c we show the E-C equilibrium coordinates for the BH systems with $N/L = 7/7, 8/7, 7/8$, and point out that the three sets, blue, green and black of states equilibrating to the CUE coordinate, equilibrating to a different value and no reaching equilibrium, identified in Fig. \ref{fig:E_C_plane_equilibrum}a have validity for these other system sizes.

The inverse participation ratios of all the states in the Fock basis for each of the BH systems with $N/L = 7/7, 8/7, 7/8$, are shown in Fig.~\ref{fig:more_ipr}a-c. We have ordered the Fock basis such that states belonging to each of the three sets, blue, green and black, will be display together. Notice how there are two orders of magnitude separating the IPR of a typical state in the blue set and a state on the black set. Pointing at the difference between ergodic states an nonergodic states, additionally ergodic states in the blue staurate the IPR averaged over the COE, dashed red line in Fig.~\ref{fig:more_ipr}a-c, ${\rm IPR_{\rm COE}} = \frac{3.0}{\mathcal{D}_{\rm BH}+1}$ see \cite{Sieberer2019}. On the other hand, states on the green set sit in between the IPR values of the blue and black sets, and thus we expect them to display a mixture of ergodic and nonergodic properties.

\bibliography{EC_plane_equilibration}

\end{document}